\documentclass{appolb}
\usepackage{epsfig}

\begin{document}
\title{New Parametrizations for the Photon Structure Function%
\thanks{Presented at the PHOTON2005 Conference,  30 August--4 September 2005,
Warsaw, Poland. I thank P. Jankowski and M. Krawczyk for the nice atmosphere in
the collaboration and PJ for his  help in preparation of  this talk. 
I also thank M.
Krawczyk and H. Abramowicz for inviting me to this very nice meeting.
This work has been partially supported by Junta de Andalucia 
(contract FQM-330),
Ministerio de educaci\'on y Ciencia (contract FPA2003-09298-c02-01) and 
the European Community's Human Potential Programme 
(contract HPRN-CT-2002-00311 EURIDICE).}%
}
\author{Fernando Cornet
\address{Departamento de F{\'\i}sica Te\'orica y del Cosmos \\
and \\
Centro Andaluz de F{\'\i}sica de Part{\'\i}culas \\   
         Universidad de Granada \\
         E-18071 Granada (Spain) }
}
\maketitle
\begin{abstract}
In the last year four new parametrizations of the Hadronic Photon Structure Function 
at Next to Leading Order have
appeared. In this talk, I briefly review the main features of three of them: the FFNS$_{CJK}$, 
CJK and AFG.
\end{abstract}
  
\section{Introduction}
The photon structure function has been  recognized as an interesting quantity 
for QCD since long ago \cite{ZERWAS,WITTEN} because it was expected that the
asymptotic point-like $Q^2$ evolution could be calculated without additional
assumptions. Unfortunately, further studies showed the need for a hadronic 
component that required extra assumptions at an input scale. However,
a good knowledge of the parton content of the photon is still needed and useful
for many phenomenological aplications. A review of the 
situation in the early days can be found in these proceedings \cite{BURAS}.

The main problem found to study the Photon Structure Function some years ago 
was the lack of experimental data \cite{KRAWCZYK,NISIUS}. 
Indeed, there were very few data and they 
covered  a very limited region in the plane $(x,Q^2)$. The situation has 
improved very much in the last years with the measurements performed by the
four LEP experiments. These measurements reduced the experimental errors
in regions of $(x,Q^2)$ that were already studied at previous experiments 
and also covered regions in this plane where there were no previous 
measurements. These has prompted the appearance in the last year of
four new parametrizations for the Photon Structure Function at LO and NLO. 
They are, in
chronological order: FFNS$_{CJK}$, CJK \cite{CJK}, AFG05 \cite{AFG} and
SAL \cite{SAL}. In this talk I will just cover the first three
parametrizations for the fourth one will be covered in the next talk 
\cite{SLOMINSKI}.

One important difference, certainly not the only one, among the new parametrizations 
is the way they deal with the heavy quark thresholds. There are three schemes to 
introduce these thresholds. 
\begin{itemize}
\item The Fixed Flavor Number Scheme (FFNS), where one considers
only the three light quarks and gluons as partons of the photon for all energy 
scales. The heavy quarks, $c$ and $b$, contribute only as external particles in the
final state produced in hard processes either in a direct production or through
the partonic content of the photon. In the calculation of the heavy quark 
contributions one keeps their mass fixed to their physical value. This scheme is
expected to give a poor description of the Photon Structure Function for 
energy scales much larger than the heavy quark masses, where one would expect the 
contributions of the heavy quarks to be similar to the ones of the light quarks. 
\item  The Zero Mass Variable Flavor Number Scheme (ZVFNS). In this scheme the 
number of active flavors as partons of the photon increases in one whenever
the energy goes through a heavy quark threshold. For then on, the heavy quark
is treated as massless in the evolution of the parton densities, just in the
same way as the light quarks are treated. This scheme is expected to solve the
problem of the FFNS scheme at large energies but, obviously, should have problems
at energies near the thresholds, where one can not neglect the heavy quark masses.
\item The Variable Flavor Number Scheme (VFNS) attempts to solve the problems of
the previous schemes. Here, one considers both contributions: the heavy quarks are 
produced in the final state taking into account their masses but also they are 
included as massless partons of the photon. In this way both energy regions
are treated properly. Unfortunately, this scheme is not free of problems either.
It is clear that there is a double counting that should be avoided introducing
some subtraction terms.
\end{itemize}

\section{The FFNS$_{CJK} NLO$ and CJK NLO Parametrizations}
It is clear from the name that the FFNS$_{CJK}$ uses the Fixed Flavor Number Scheme,
while the CJK parametrization is using the VFNS. Detailed expressions for the 
Photon Structure Function, $F_2^\gamma (x,Q^2)$, for both parametrizations, involving a 
description of the way the subtraction terms are chosen in the CJK parametrization 
can be found in Ref. \cite{CJK}. In addition, the CJK parametrization uses the
ACOT($\chi$) scheme. The idea is to enforce that the heavy quark distribution
functions vanish for $W = 2 m_h$ (and below), where $W$ is the invariant mass of the hadronic
final state. This is achieved substituting the $x$ variable 
by $\chi_h = x(1+4m_h^2/Q^2)$ in the heavy quark densities. In this way
$\chi \to 1$ and $q_h(\chi, Q^2) \to 0$ for $W^2=(1-x)Q^2/x \to 4 m_h^2$.

Both parametrizations are written as a function
of the quark and gluon distribution functions that obey an inhomogeneous DGLAP set
of equations. In order to solve these equations we introduce the same input for both
parametrizations at $Q_0^2 = 0.765 \;$ GeV$^2$, based on Vector Meson Dominance (VMD):
\begin{equation}
f^\gamma(x,Q_0^2) = \sum_{V}\frac{4\pi \alpha}{\hat f^2_{V}}f^{V}(x,Q_0^2), 
\end{equation}
with the sum running over all light vector mesons (V) into which the photon can 
fluctuate. The parameters $\hat f^2_{V}$ can be extracted from the experimental
data on $\Gamma(V\to e^+e^-)$ width. In practice we take into account the  
$\rho^0$ meson while the contributions from the other mesons are accounted for 
via a parameter $\kappa$
\begin{equation}
f^\gamma(x,Q_0^2) = \kappa\frac{4\pi \alpha}{\hat f^2_{\rho}}f^{\rho}(x,Q_0^2),
\label{vmdfor}
\end{equation}
which is left as a free parameter in the fits. The scale $Q_0$ has been fixed to
this value because it is the one that allows a better fit to the experimental data.

For the $\rho$ meson we assume the following form for the valence quark and gluon 
distributions
\begin{eqnarray}
xv^{\rho}(x,Q_0^2) &=& N_v x^{\alpha}(1-x)^{\beta}, \label{input1} \\
xG^{\rho}(x,Q_0^2) &=& \tilde N_G xv^{\rho}(x,Q_0^2)=
N_G x^{\alpha}(1-x)^{\beta}, \nonumber
\end{eqnarray}
where $N_v$, $N_G$, $\alpha$ and $\beta$ are free parameters. The sea quark distribution
is assumed to vanish at this scale.
This is similar to
what was done in the GRV parametrization \cite{GRV}, but there the authors fixed the 
values of the parameters $\alpha$ and $\beta$ to the ones they had previously obtained 
for the pion distribution functions. Since there are more data available now, we 
prefer to leave these parameters as free parameters in the fit.  

We have included in the fit all the available data in year 2004 except 
the DELPHI LEP2 data because 
they present three sets of mutually inconsistent data. The total number of points used in the 
fit is $192$ covering a kinematical range of $0.001  \leq x \leq 0.65$ and 
$1.3 \; <$GeV$^2 \leq Q^2 \leq 780 \; $GeV$^2$. The results of the fit are shown in Table 1.
Introducing the DELPHI LEP2 data in the fit the  $\chi^2/_{DOF}$ increases to
$1.50$ (TWOGAM), $1.54$ (PHOJET) or $1.66$ (PYTHIA), depending on the MonteCarlo used to
analise the data.

\begin{table}[h]
\begin{center}
\renewcommand{\arraystretch}{1.5}
\begin{tabular}{|c|@{} p{0.1cm} @{}|c|c|@{} p{0.1cm} @{}|c|c|c|}
\hline
 NLO models         && $\chi^2$ & $\chi^2/_{DOF}$ && $\kappa$ & $\alpha$ & $\beta$ \\
\hline
\hline
 FFNS$_{CJK}$ && 243.3 & 1.29 && 2.288$^{+0.108}_{-0.096}$ & 0.502$^{+0.071}_{-0.066}$ & 0.690$^{+0.282}_{-0.252}$ \\
\hline
 CJK           && 256.8 & 1.37 && 2.662$^{+0.108}_{-0.099}$ & 0.496$^{+0.063}_{-0.057}$ & 1.013$^{+0.284}_{-0.255}$  \\
\hline
\end{tabular}
\caption{\small The $\chi^2$ and parameters of the final fits for 192 data 
points for FFNS$_{CJK}$ NLO and CJK NLO models with assumed $Q_0^2 = 0.765$ 
GeV$^2$. The $\alpha$, $\beta$ and $\kappa$ errors are obtained from \textsc{Minos}
requiring $\Delta \chi^2 = 1$.}
\label{tparam}
\end{center}
\end{table} 

A comparison of the CJK predictions with the recent L3 data for $Q^2 = 12.4 \; $GeV$^2$ 
and $Q^2 = 16.7 \; $GeV$^2$, not included in the fit because they have been published 
after the fit was performed \cite{L3,KIENZLE} is shown in Fig. 1. Data for similar $Q^2$
from CELLO \cite{CELLO}, DELPHI \cite{DELPHI}, OPAL \cite{OPAL} and TOPAZ \cite{TOPAZ}
are also included. We see that the CJK parametrization provides a good description
of the data, even though not all the data sets are fully compatible among each other.

\begin{figure}
\begin{center}
\epsfxsize=13cm
\epsfbox{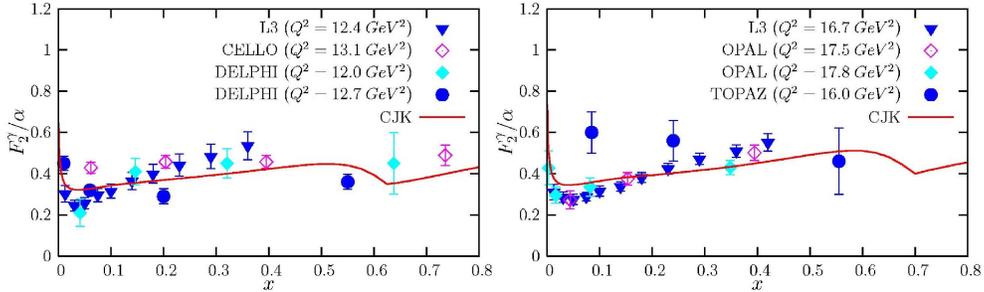}
\caption{Comparison of the CJK NLO prediction with various sets of data, including
the new L3 data not included in the fit. The kink observed in the curve at large $x$
is the charm quark threshold.}
\end{center}
\end{figure}

\section{The AFG NLO Parametrization}

The third parametrization I will briefly review here has been performed by Aurenche, Fontannaz
and Guillet \cite{AFG}. It is an update of a previous parametrization obtained by the 
same authors \cite{AFGold}. This parametrization uses the ZVFNS with $N_f = 5$, however
they keep terms $O(m_h^2/Q ^2)$ in the direct contribution in order to have a smooth threshold
behaviour. At $Q^2 = Q_0^2$ the structure function is given by:
\begin{equation}
{F_2^\gamma (x,Q_0^2) \over x} = C_\gamma (x) + 
   \sum_{f=1}^{N_f} [e_f^2 (q_f^{NP} (Q_0^2) + \bar q_f^{NP} (Q_0^2) - C_{\gamma,c}^f],
\end{equation}
where $C_\gamma (x)$ is the direct contribution and $C_{\gamma , c}^f$ is given
by the ``hand-bag'' diagram. The non-perturbative input is also based on VMD, identifying
the form of the parton distributions for the $\rho$ meson with the ones for the pion
obtained in Ref. \cite{AURENCHE}, but leaving a normalization factor, $C_{np}$, 
as a free constant: 
\begin{equation}
\begin{array}{c}
x u^\gamma_{valence} = C_{np} \alpha {4 \over 9} x u^\pi_{valence} 
     =  C_{np} \alpha {4 \over 9} {1 \over B(p_2,1+p_3)} x^{p_2} (1-x)^{p_3} \vspace*{0.2cm}\\
x u^\gamma_{sea} = C_{np} \alpha {2 \over 3} x u^\pi_{sea} 
     = C_{np} \alpha {2 \over 3} C_s (1-x)^{p_8} \vspace*{0.2cm} \\
x G^\gamma = C_{np} {2 \over 3} x g^\pi = C_{np} {2 \over 3} C_g (1-x)^{p_{10}},
\end{array}
\end{equation}
where $p_2 = 0.48$, $p_3 = 0.85$, $p_8 = 7.5$, $p_{10} = 1.9$, $C_s = 1.2$, 
$C_g = 0.447(1+p_{10})$ and $B(x,y)$ is the beta function. In summary, there are only two 
free parameters: the input scale $Q_0$
and the normalization factor $C_{np}$.

The values of the two free parameters are obtained performing a fit to all the LEP experimental
data. The best fit, with a $\chi^2/DOF = 1.03$, gives $Q_0 = 0.7 \; $GeV$^2$ (very similar to 
the one used in the FFNS$_{CJK}$ and CJK parametrizations) and $C_{np} = 0.78$. Aurenche, Fontannaz
and Guillet have also performed independent fits for each one of the four LEP experiments.
The result is summarized in Fig. 2, where one can see that the best fits from the DELPHI 
experiment give very different values of the parameters compared with the ones obtained from the
global fit as well as from each one of the other three experiments.
\begin{figure}
\begin{center}
\epsfxsize=6cm
\epsfbox{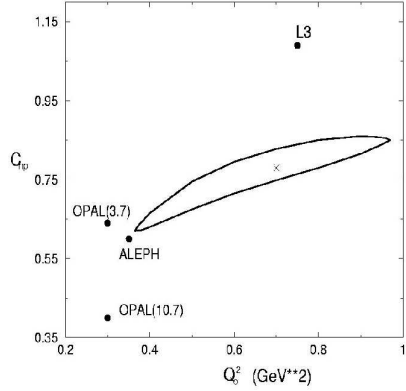}
\caption{Figure from Ref. \cite{AFG} where it is presented the $\Delta \chi^2 = 1$ contour in the
($Q_0^2,C_{np}$) plane as well as the individual best fits for each LEP experiment. The point
for DELPHI(12.7) is outside the figure, while for DEPHI (3.7) no minimum is found for 
$Q_0^2 < 1.9 \; $GeV$^2$.}
\end{center}
\end{figure}

The authors have also explored two other parametrizations allowing for a harder gluon component
modifying the value of $p_{10}$to $p_{10} = 1.0$ or a softer gluon component with
$p_{10} = 4.0$.

\section{Web Pages}
Instead of a summary I will finish just refering the interested reader to the web pages where he can find
FORTRAN routines with these parametrizations:
\begin{itemize}
\item FFNS$_{CJK}$ and CJK: http://www.fuw.edu.pl/$\sim$pjank/param.html
\item AFG05: http://www.lapp.in2p3.fr/lapth/PHOX\underline{~}FAMILY/main.html
\end{itemize}


\end{document}